\documentclass[reprint, superscriptaddress, twocolumn, amsmath, amssymb, aps, prb]{revtex4-2}
\usepackage{graphicx}
\usepackage{dcolumn}
\usepackage{bm}
\usepackage{placeins}
\usepackage{appendix}
\usepackage{upgreek}

\usepackage{verbatim}
\usepackage[usenames,dvipsnames]{xcolor}
 \usepackage{amsmath}
 \usepackage{amsfonts}
 \usepackage{amssymb}
 \usepackage[colorlinks=true,citecolor=blue,linkcolor=red]{hyperref}

 \usepackage{bbold}     
 \usepackage[makeroom]{cancel}  
 \usepackage{multirow}    
 \usepackage[normalem]{ulem}        
 \usepackage{array}
 \usepackage{pdfpages}
\makeatletter
 \AtBeginDocument{\let\LS@rot\@undefined}
 \makeatother

\begin{document}


\title{Gate-tunable superconductor-semiconductor coupling and its impact on zero-energy modes in PbTe-Pb nanowires}


\author{Ruidong Li}
\email{equal contribution}
\affiliation{College of Semiconductors, Southern University of Science and Technology, Shenzhen 518055, China}
\affiliation{State Key Laboratory of Low Dimensional Quantum Physics, Department of Physics, Tsinghua University, Beijing 100084, China}

\author{Shuai Yang}
\email{equal contribution}
\affiliation{State Key Laboratory of Semiconductor Physics and Chip Technologies, Institute of Semiconductors, Chinese Academy of Sciences, Beijing 100083, China}

\author{Wenyu Song}
\email{equal contribution}
\affiliation{State Key Laboratory of Low Dimensional Quantum Physics, Department of Physics, Tsinghua University, Beijing 100084, China}

\author{Zeyu Yan}
\affiliation{State Key Laboratory of Low Dimensional Quantum Physics, Department of Physics, Tsinghua University, Beijing 100084, China}

\author{Yuhao Wang}
\affiliation{State Key Laboratory of Low Dimensional Quantum Physics, Department of Physics, Tsinghua University, Beijing 100084, China}

\author{Zehao Yu}
\affiliation{State Key Laboratory of Low Dimensional Quantum Physics, Department of Physics, Tsinghua University, Beijing 100084, China}

\author{Li Chen}
\affiliation{School of Physics and Optoelectronic Engineering, Guangdong University of Technology, Guangzhou 510006, China}

\author{Zixun Xia}
\affiliation{State Key Laboratory of Low Dimensional Quantum Physics, Department of Physics, Tsinghua University, Beijing 100084, China}

\author{Zonglin Li}
\affiliation{State Key Laboratory of Low Dimensional Quantum Physics, Department of Physics, Tsinghua University, Beijing 100084, China}

\author{Shan Zhang}
\affiliation{State Key Laboratory of Low Dimensional Quantum Physics, Department of Physics, Tsinghua University, Beijing 100084, China}

\author{Jiaye Xu}
\affiliation{State Key Laboratory of Low Dimensional Quantum Physics, Department of Physics, Tsinghua University, Beijing 100084, China}

\author{Lining Yang}
\affiliation{State Key Laboratory of Low Dimensional Quantum Physics, Department of Physics, Tsinghua University, Beijing 100084, China}

\author{Weizhao Wang}
\affiliation{State Key Laboratory of Low Dimensional Quantum Physics, Department of Physics, Tsinghua University, Beijing 100084, China}

\author{Xiao Feng}
\affiliation{State Key Laboratory of Low Dimensional Quantum Physics, Department of Physics, Tsinghua University, Beijing 100084, China}
\affiliation{Beijing Academy of Quantum Information Sciences, Beijing 100193, China}
\affiliation{Frontier Science Center for Quantum Information, Beijing 100084, China}
\affiliation{Hefei National Laboratory, Hefei 230088, China}

\author{Tiantian Wang}
\affiliation{Beijing Academy of Quantum Information Sciences, Beijing 100193, China}
\affiliation{Hefei National Laboratory, Hefei 230088, China}

\author{Yunyi Zang}
\affiliation{Beijing Academy of Quantum Information Sciences, Beijing 100193, China}
\affiliation{Hefei National Laboratory, Hefei 230088, China}

\author{Lin Li}
\affiliation{Beijing Academy of Quantum Information Sciences, Beijing 100193, China}

\author{Runan Shang}
\affiliation{Beijing Academy of Quantum Information Sciences, Beijing 100193, China}
\affiliation{Hefei National Laboratory, Hefei 230088, China}

\author{Dong E. Liu}
\affiliation{State Key Laboratory of Low Dimensional Quantum Physics, Department of Physics, Tsinghua University, Beijing 100084, China}
\affiliation{Beijing Academy of Quantum Information Sciences, Beijing 100193, China}
\affiliation{Frontier Science Center for Quantum Information, Beijing 100084, China}
\affiliation{Hefei National Laboratory, Hefei 230088, China}

\author{Qi-Kun Xue}
\affiliation{State Key Laboratory of Low Dimensional Quantum Physics, Department of Physics, Tsinghua University, Beijing 100084, China}
\affiliation{Beijing Academy of Quantum Information Sciences, Beijing 100193, China}
\affiliation{Frontier Science Center for Quantum Information, Beijing 100084, China}
\affiliation{Hefei National Laboratory, Hefei 230088, China}
\affiliation{Department of Physics, Southern University of Science and Technology, Shenzhen 518055, China}

\author{Ke He}
\email{kehe@tsinghua.edu.cn}
\affiliation{State Key Laboratory of Low Dimensional Quantum Physics, Department of Physics, Tsinghua University, Beijing 100084, China}
\affiliation{Beijing Academy of Quantum Information Sciences, Beijing 100193, China}
\affiliation{Frontier Science Center for Quantum Information, Beijing 100084, China}
\affiliation{Hefei National Laboratory, Hefei 230088, China}

\author{Hao Zhang}
\email{zhanghao@sustech.edu.cn}
\affiliation{College of Semiconductors, Southern University of Science and Technology, Shenzhen 518055, China}

\begin{abstract}

Superconductor-semiconductor coupling in hybrid nanowires plays a pivotal role in Majorana devices. Here, we demonstrate this tunable coupling in PbTe-Pb nanowires and its impact on the induced gap and zero-bias peaks (ZBPs). In the strong coupling regime, the device reveals a clean superconducting gap, where robust and non-robust ZBPs are observed in distinct magnetic field ranges. Conversely, increasing the gate voltage drives the system toward the weak coupling regime, where the gap becomes narrow and soft. Consequently, robust ZBPs disappear while non-robust ones persist. Our results establish a direct link between tunable coupling and robust ZBPs, providing guidelines on engineering Majorana zero modes and topological qubits.  

\end{abstract}

\maketitle

Majorana zero modes in hybrid superconductor-semiconductor nanowires are an elegant theoretical concept \cite{Lutchyn2010, Oreg2010}. After a decade of study, their experimental realization, however, remains challenging \cite{Mourik, Deng2016, Albrecht, Nichele2017, Gul2018, Song2022, WangZhaoyu, Delft_Kitaev, MS_2023, MS_2025, NextSteps, Prada2020, Leo_perspective}. The devices often exhibit more complex phenomena than initially expected \cite{Patrick_Lee_disorder_2012, Prada2012, Loss2013ZBP, GoodBadUgly, DasSarma_estimate, DasSarma2021Disorder, Tudor2021Disorder, Loss_Andreev_band}. Theoretically, a superconductor can induce a superconducting gap ($\Delta$) in a semiconductor nanowire through the proximity effect. Applying a magnetic field ($B$) beyond a critical value is expected to drive the nanowire into a topological superconducting phase, where Majorana zero modes can emerge. This topological criterion can be expressed by the condition: $E_{\text{Z}}>\sqrt{\Delta^2+\mu^2}$, where $E_{\text{Z}}=\frac{1}{2}g\mu_{\text{B}}B$ is the Zeeman energy, $\mu$ is the electro-chemical potential, $g$ is the effective Land\'{e} $g$-factor, and $\mu_{\text{B}}$ is the Bohr magneton. The superconductor-semiconductor coupling is crucial as it influences all three terms in this inequality. For example, the coupling strength sets the size of the induced gap $\Delta$, while the gap quality (hard vs soft) depends on the coupling uniformity \cite{Takei2013, Chang2015, Gul2017}. The effective $g$-factor arises from hybridization, namely, the mixing of $g$-factors between the superconductor and semiconductor. The coupling strength governs the hybridization details \cite{LutchynSchrodinger, Flensberg_SP,DasSarma_SP, Loss_metalization,CPH2018effectiveg,  Michiel2018, Nick_NatureComm} and can also modulate the amount of charge transfer to the semiconductor, indirectly affecting $\mu$. Another important physical quantity implicit in this framework is the spin-orbit interaction, which is also modulated by charge transfer and interface coupling \cite{Jouri2019}. 

Although controlling this coupling is paramount, electrostatic gating remains the primary method for its in situ modulation once the device is fabricated. Previous studies have demonstrated gate-tunable coupling in InSb-Al \cite{Michiel2018, Nick_NatureComm} and InAs-Al \cite{CPH2018effectiveg, MS_2023} hybrids. However, a direct correlation with Majorana signatures has yet to be clearly established. Here, we report tunable coupling in PbTe-Pb hybrid nanowires and, more importantly, its impact on the emergence of robust and non-robust zero-bias peaks (ZBPs). By establishing the direct link between coupling strength and robust ZBPs, our results can guide future experiments on PbTe-Pb Majorana devices.  

\begin{figure}[htb]
\includegraphics[width=\columnwidth]{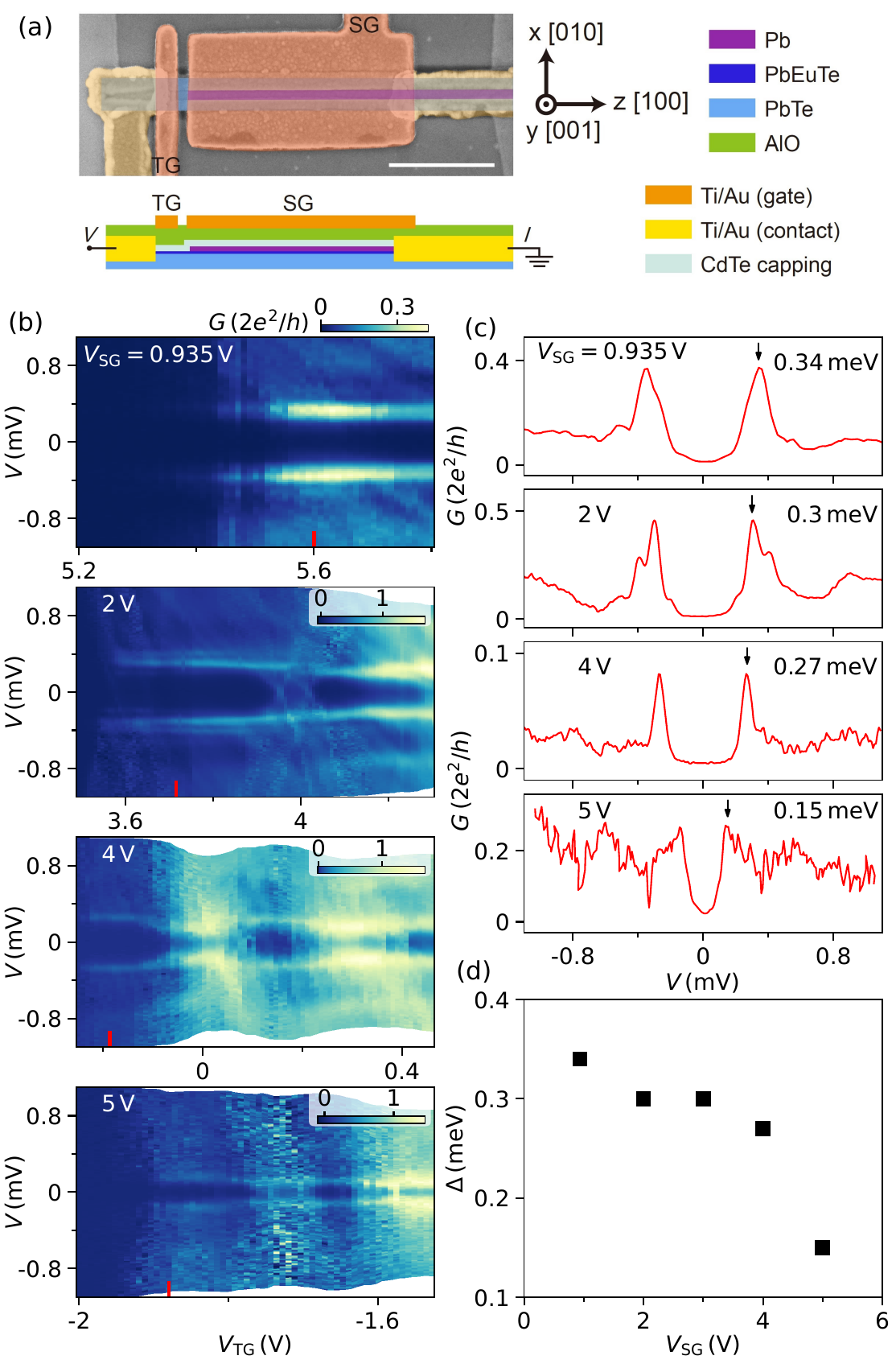}
\centering
\caption{Gate-tunable coupling. (a) False-colored SEM and schematic of the PbTe-Pb device. (b) $G$ vs $V$ and $V_{\text{TG}}$ for $V_{\text{SG}}$ = 0.935 V, 2.0 V, 4.0 V, and 5.0 V, respectively. (c) Line cuts from (b), labeled by red bars. (d) Extracted gap size as a function of $V_{\text{SG}}$. $B$ = 0 T.}
\label{fig1}
\end{figure}

\begin{figure}[htb]
\includegraphics[width=\columnwidth]{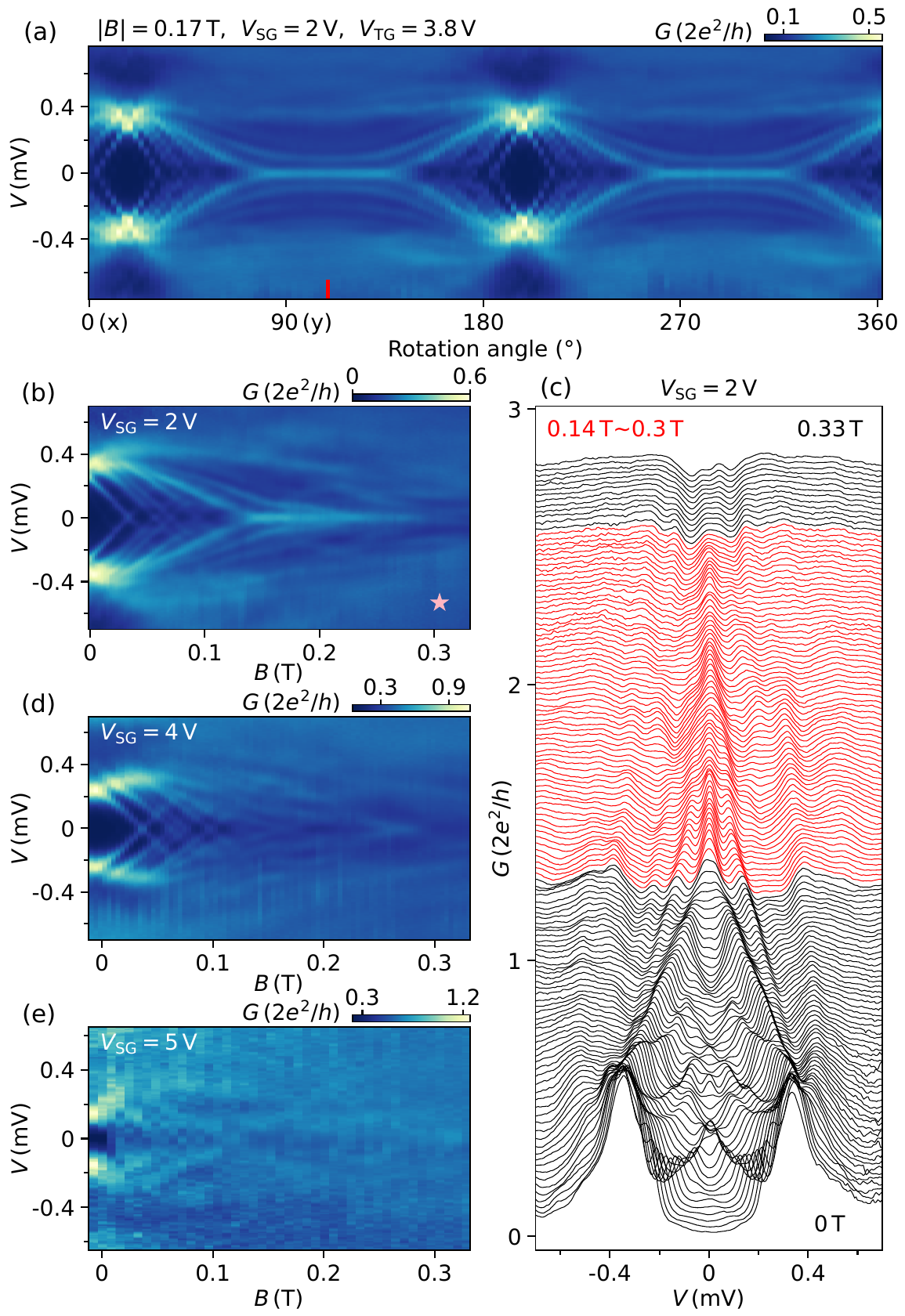}
\centering
\caption{ZBPs in different coupling regimes. (a) $B$ rotation in the $xy$ plane. $|B|$ = 0.17 T. (b) $G$ vs $V$ and $B$. $V_{\text{TG}}$ = 3.8 V, $V_{\text{SG}}$ = 2.0 V for (a-b). (c) Waterfall plot of (b). Vertical offset, 0.02$\times2e^2/h$ for clarity. (d-e) $G$ vs $V$ and $B$ for $V_{\text{SG}}$ = 4.0 V and 5.0 V, respectively. $V_{\text{TG}}$ = 1.51 V and -0.855 V. $B$ direction is 109$^{\circ}$ from the $x$ axis (see the red bar in (a)).    }
\label{fig2}
\end{figure}

\begin{figure*}[htb]
\includegraphics[width=\textwidth]{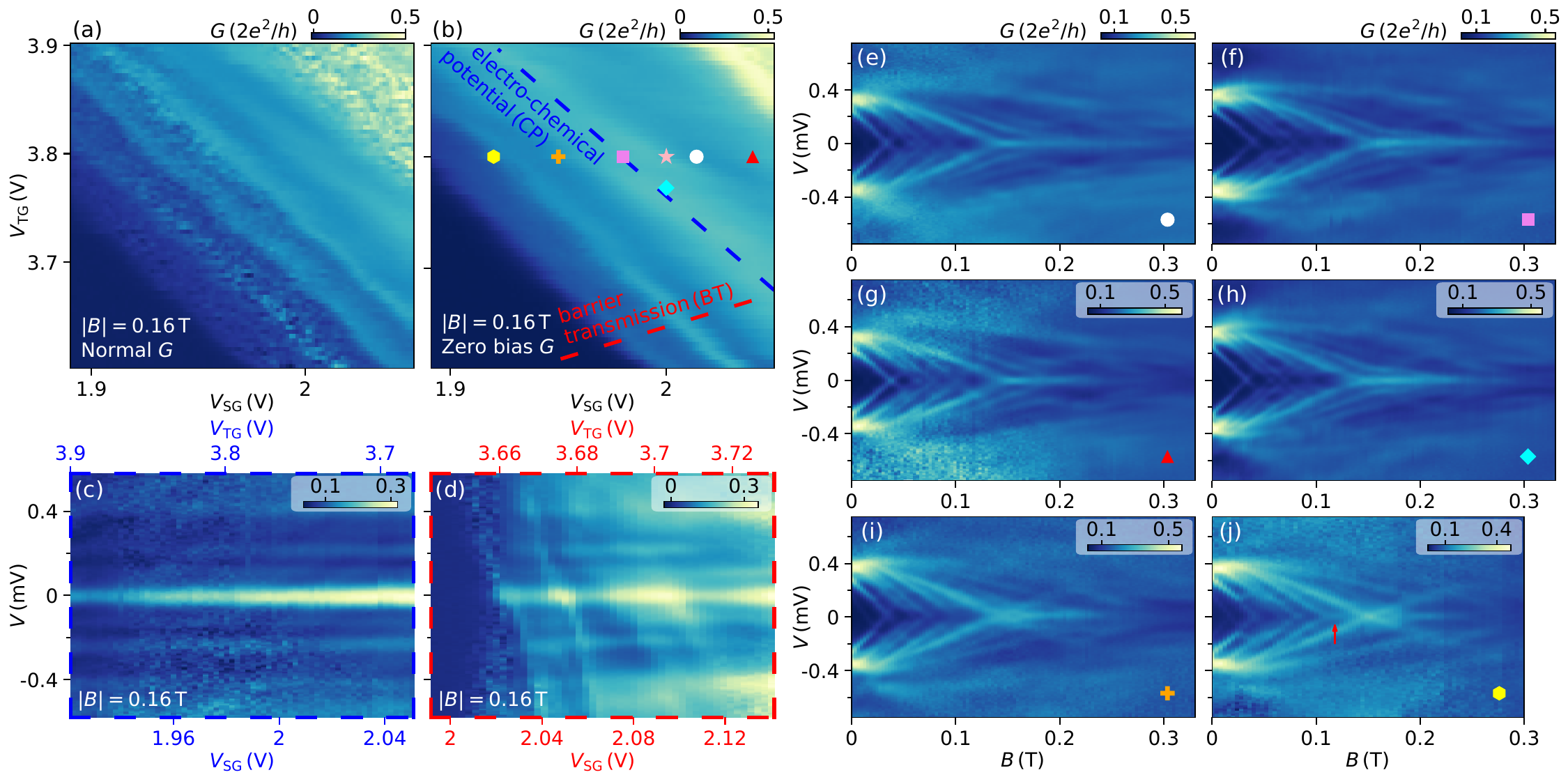}
\centering
\caption{Robust ZBPs in the strong coupling regime. (a-b) $G$ vs $V_{\text{TG}}$ and $V_{\text{SG}}$ for normal state (a) and zero bias (b). $B$ = 0.16 T. (c) Bias scan along the blue dashed line in (b). (d) Bias scan along red dashed line in (b) (its actual gate range exceeds the border of (b)). (e-j) Six $B$ scans with their gate voltages marked in (b).   }
\label{fig3}
\end{figure*}

Figure 1(a) shows a scanning electron micrograph (SEM) of the PbTe-Pb device and its schematic. The right panel defines the coordinate axes and crystal orientation. For an overview of previous studies on PbTe devices, we refer to Refs. \cite{CaoZhanPbTe, Jiangyuying, Erik_PbTe_SAG, PbTe_AB, Fabrizio_PbTe, Zitong_JJ, Wenyu_QPC, Yichun_Gap, Yuhao_QPC, Ruidong_PJJ, Vlad_PbTe, PbTe_In, Yichun_SQUID, Yuhao_degeneracy, Wenyu_Disorder, Quantized_Andreev, Zonglin_Anisotropy, Zhangshan_ZBP, PbTe_SC_theory}. The width of the Pb film (false-colored violet) is approximately one-third that of the PbTe nanowire (false-colored blue). The device adopts a standard normal-superconductor (``NS'') geometry with two top gates: the tunneling gate (TG) and the ``super gate'' (SG). The TG tunes the barrier transmission, while the SG tunes the proximitized region. Transport measurements were performed in a two-terminal configuration in a dilution refrigerator at a base temperature below 50 mK. For details regarding device growth and fabrication, we refer to the Supplemental Material (SM).

Figure 1(b) shows the differential conductance, $G\equiv dI/dV$, as a function of $V_{\text{TG}}$ and bias voltage ($V$). $I$ is the current. Each panel was measured at a fixed $V_{\text{SG}}$. In the tunneling regime, a superconducting gap is observed, see line cuts in Fig. 1(c). The black arrows indicate the positions of the coherence peaks, which we define as gap edges. As $V_{\text{SG}}$ increases, the gap size ($\Delta$) decreases from 0.34 meV to 0.30 meV, 0.27 meV, and finally 0.15 meV. Figure 1(d) summarizes this downward trend. Moreover, the gap becomes increasingly soft, i.e. with increasing subgap conductance, as $V_{\text{SG}}$ becomes more positive. This evolution is compatible with the gate tunablity of the superconductor-semiconductor coupling, similar to that in III-V nanowires \cite{Michiel2018, Nick_NatureComm, CPH2018effectiveg}. A more positive $V_{\text{SG}}$ attracts the electron wavefunction deeper into the PbTe nanowire, reducing its weight in the Pb shell. Consequently, the coupling weakens and the induced gap decreases \cite{LutchynSchrodinger, Flensberg_SP,DasSarma_SP}. The gap becomes softer because the electrons are spatially further away from the Pb film. The modulation is dramatic, indicating that the proximitized PbTe region is highly gate-tunable despite the screening of its large dielectric constant ($\sim$ 1350). 

Guided by the evolution of the induced gap, we define the super-gate range $V_{\text{SG}} \sim$1.0-2.0 V as the strong coupling regime, whereas $V_{\text{SG}} \sim$4.0-5.0 V corresponds to the weak-coupling regime. These designations are relative to the current device's tunable range. In fact, due to the presence of the 4-nm-thick Pb$_{0.97}$Eu$_{0.03}$Te interlayer (dark blue in Fig. 1(a)) grown between PbTe and Pb, even the induced gap in the strong coupling regime is still significantly smaller than the bulk gap of Pb ($\sim$ 1 meV). Thus, "strong" coupling observed here would likely be categorized as "intermediate" compared to the PbTe-Pb devices lacking such an interlayer \cite{Yichun_Gap}.

Next, we investigate the strong coupling regime by setting $V_{\text{SG}}$ = 2 V, as shown in Figs. 2(a-b). A robut ZBP is observed at finite $B$'s. In Fig. 2(a), we fix $|B|$ at 0.17 T and rotate its direction in the $xy$ plane. The coordinate axis is sketched in Fig. 1(a) (and also Fig. S1(c)). The $y$-axis is out-of-plane and the $x$-axis is in-plane and perpendicular to the nanowire. The rotating angle is measured relative to the $x$ axis. The ZBP remains robust within an angular range from 79$^{\circ}$ to 139$^{\circ}$. Accordingly, the $B$ direction was fixed at 109$^{\circ}$ for the rest of the measurement unless specified. For rotations in other planes, see Fig. S1 in the SM.

Figure 2(b) illustrates the $B$ scan where the robust ZBP persists from 0.14 T to 0.30 T. Its waterfall plot is shown in Fig. 2(c). The peak height is well below $2e^2/h$, possibly due to thermal broadening and gap softening at finite fields \cite{Nichele2017, LiuDissipation, Law2009, Flensberg2010}. Above 0.30 T, the ZBP splits as the gap starts to close. Below 0.14 T, several subgap states originating near the gap edge cross zero energy with increasing $B$. These sharp crossings result in non-robust (fine-tuned) ZBPs. The robust and non-robust ZBPs appear in different $B$ ranges, provides a stark contrast. This phenomenon is consistent with a theoretical simulation \cite{Prada2012}, in which a quantum dot is formed in the barrier region. In such scenarios, several dot-induced Andreev bound states \cite{Mason2011, Silvano2014} can lead to fine-tuned ZBPs prior to the topological transition and the subsequent emergence of a robust Majorana mode.

From the slopes of the $B$ dependent subgap states, we estimate the effective $g$-factor to range from 110 to 270. These values are an order of magnitude larger than those reported in tunneling spectroscopy experiments on InSb and InAs (typically $<$10) \cite{Deng2016, Gul2018, Song2022, WangZhaoyu, MS_2023}. Consequently, the onset field for the robust ZBPs ($\sim$ 0.14 T) is significantly lower than the $\sim$1 T typically required in III-V systems. This  $g$-factor enhancement is likely driven by orbital effects, as $B$ is nearly out-of-plane (tilted only 19$^{\circ}$ from the $y$ axis). Studies \cite{Wimmer2017Orbital, full_shell, full_shell_erratum, Zhangshan_ZBP} have shown that orbital contributions to the $g$-factor can lower the critical field for the emergence of zero modes. Furthermore, this orbital effect is sensitive to $B$ orientation, leading to significant $g$-factor anisotropy \cite{Zonglin_Anisotropy, Zhangshan_ZBP}. Figure 2(a) can be interpreted within this framework: the variation in the $g$-factor during rotating effectively modulates the Zeeman energy, similar to that in Fig. 2(b).

We then drove the device into the weak coupling regime by setting $V_{\text{SG}}$ to 4 V, as shown in Fig. 2(d). The zero-field gap is reduced and notably, no robust ZBPs are observed at finite fields. In contrast, dot-induced non-robust ZBPs persist. For its waterfall plot, see Fig. S2 in the SM. These dot-induced ZBPs have been well-studied \cite{Patrick_Lee_disorder_2012, Prada2012, Loss2013ZBP, GoodBadUgly, DasSarma_estimate, DasSarma2021Disorder, Tudor2021Disorder, Loss_Andreev_band}. In a Majorana interpretation, an increase in $V_{\text{SG}}$ elevates $\mu$, thereby raising the onset field for the topological transition. When this onset field exceeds the critical field of the induced superconductivity, Majoranas and robust ZBPs do not emerge. We emphasize that while non-robust ZBPs are clearly not topological, robust ones can be consistent with a Majorana scenario but do not constitute sufficient evidence. It would be interesting to explore possible alternative explanations in future studies. For even more positive $V_{\text{SG}}$ (Fig. 2(e)), the gap becomes narrower and softer, indicating weakened induced superconductivity. Both robust ZBPs and fine-tuned dot states disappear in this regime.

In Fig. 3, we perform an in-depth analysis of the robust ZBPs in the strong coupling regime. Figure 3(a) shows the conductance map in ($V_{\text{TG}}$, $V_{\text{SG}}$) at $B$ = 0.16 T. $V_{\text{SG}}$ varies between 1.89 V and 2.05 V, within the strong coupling regime. The total bias voltage $V_{\text{bias}}$ \cite{bias} is fixed at 0.985 mV for (a) and 0 mV for (b), respectively. Figure 3(a) thus represents the normal-state $G$, while Fig. 3(b) depicts the zero-bias $G$. The regions of constant conductance exhibit a tilt, indicating that both $V_{\text{TG}}$ and $V_{\text{SG}}$ can tune the barrier region. Dot states formed near this region are revealed as resonances. To compensate for the gate cross-talk, we define the blue dashed line as the gate of electro-chemical potential ($V_{\text{CP}}$) \cite{Deng2016}. By scanning along this line, the barrier transmission is kept constant while $\mu$ is tuned (see Fig. 3(c)). The ZBP remains robust by sweeping $\mu$. We define the red dashed line in Fig. 3(b) as $V_{\text{BT}}$, which mainly tunes the barrier transmission and dot states. Figure 3(d) shows the sweep along $V_{\text{BT}}$, also revealing a robust ZBP until the pinch-off regime. Notably, $V_{\text{BT}}$ tunes several dot states across zero energy, and the ZBP remains non-split throughout these crossings. This robustness with respect to dot states has been theoretically investigated \cite{Prada2017, Clarke2017} as a possible indicator of the non-locality of Majoranas, or alternatively, as a result of asymmetric coupling to quasi-Majoranas \cite{WimmerQuasi}.

We then performed several $B$ scans, as shown in Figs. 3(e-j), with their corresponding gate configurations marked in Fig. 3(b). Their waterfall plots are provided in Fig. S3. The gate setting for Fig. 2(b) is also labeled in Fig. 3(b) by a pink star. All scans reveal robust ZBPs, indicating their prevalence across the whole explored parameter space in Fig. 3(b), excluding the pinch-off regime (lower-left corner).  All robust ZBPs in Figs. 3(e-j) and Figs. 2(a-b) emerge from the coalescence of two subgap levels, a feature reminiscent of gap closing. Gap-reopening is, however, less discernible. Instead, we frequently observe a crossover feature from a second subgap level. This crossover is equivalent to the reopening feature when the nanowire length is finite \cite{Alicea_Finite}. In Figs. 3(i-j), the reopening feature is more discernible. Theoretical simulations \cite{Das_Sarma_to_close} have shown that the visibility of gap closing and reopening in transport measurement is sensitive to the specific wavefunction distribution near the barrier region.

Prior to the gap-closing, non-robust ZBPs caused by dot states are prevalent and vary with gate voltages, highlighting their fine-tuned nature. Notably, a dot state may occasionally mimic a robust ZBP in a field sweep, as indicated by the red arrow in Fig. 3(j). However, this feature is fundamentally fine-tuned, as it only appears at a specific gate setting. Similar phenomena have been predicted theoretically \cite{Liu2017, Loss2018ABS,Tudor2018}. For additional data in the strong coupling regime, we refer to Fig. S4. 

\begin{figure}[htb]
\includegraphics[width=\columnwidth]{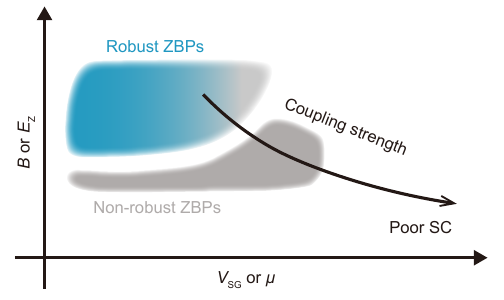}
\centering
\caption{A sketch of ZBP evolution in the parameter space.   }
\label{fig4}
\end{figure}

The robust ZBPs in the strong coupling regime remain pinned at zero bias (non-split) over a substantial range of $B$, $V_{\text{SG}}$, and $V_{\text{TG}}$, demonstrating their robustness against variations in Zeeman energy, electrochemical potential, barrier transparency, and accidental dot states. This robustness is notably absent in the weak coupling regime ($V_{\text{SG}} \sim$4-5 V).  Figure S5 shows additional $B$ scans in this regime. To illustrate the gradual evolution between these two regimes, Figure S6 displays sweeps that transition from strong to weak coupling, where the robust ZBP gradually degrades, splits, and eventually vanishes as $V_{\text{SG}}$ increases.

Figure 4 provides a schematic summary of our observations. Robust ZBPs are prevalent in the low-$\mu$ regime, spanning a $V_{\text{SG}}$ range exceeding 1 V (the blue region). Non-robust ZBPs appear at lower fields (the grey region). At higher $V_{\text{SG}}$'s, the coupling strength weakens—as evidenced by a smaller and softer induced gap—leading to the disappearance of robust ZBPs while fine-tuned ones persist. The rightmost region corresponds to poor induced superconductivity (poor SC), where both types of ZBPs disappear. The results so far are from one device. In Fig. S7, we show a second device exhibiting tunable coupling. Future efforts should demonstrate the reproducibility of the link between tunable coupling and robust ZBPs in more devices with reduced device disorder.

In conclusion, we have demonstrated the significant tunability of superconductor-semiconductor coupling in PbTe-Pb hybrid nanowires. The presence of PbEuTe interlayer enables the observation of robust ZBPs \cite{Yichun_Gap}. By increasing the super-gate voltage, the coupling strength can be tuned from strong to weak, resulting in a systematically smaller and softer gap. In the strong coupling regime, robust ZBPs emerge alongside features consistent with gap closing and reopening. Dot states coexist and cause fine-tuned ZBPs. As the coupling weakens, robust ZBPs degrade and eventually vanish, whereas fine-tuned ZBPs originating from quantum dots remain. While these observations are consistent with an interpretation involving Majorana zero modes, they do not constitute a definitive proof, as local tunneling spectroscopy is a necessary but insufficient diagnostic. Nevertheless, these results establish a direct link between tunable coupling and robust ZBPs, shedding light on future device optimizations toward conclusive evidence of Majoranas.

\section{Acknowledgment} 

This work is supported by the by National Natural Science Foundation of China (Grant Nos. 92565302 and 12504191) and Quantum Science
and Technology-National Science and Technology Major Project (2021ZD0302400). S.Y. and W. S. acknowledges the China Postdoctoral Science Foundation (Grant Nos. 2024M751610, BX20250167 and 2025M783399) and Postdoctoral Fellowship Program of China Postdoctoral Science Foundation (Grant No. GZC20231368).  

\section{Data Availability}

Raw data and processing codes within this paper are available at \cite{rawdata}.

\bibliography{mybibfile}

\newpage

\onecolumngrid

\newpage
\includepdf[pages=1]{PbTe_Tunable_SC_SM.pdf}
\includepdf[pages=2]{PbTe_Tunable_SC_SM.pdf}
\includepdf[pages=3]{PbTe_Tunable_SC_SM.pdf}
\includepdf[pages=4]{PbTe_Tunable_SC_SM.pdf}
\includepdf[pages=5]{PbTe_Tunable_SC_SM.pdf}
\includepdf[pages=6]{PbTe_Tunable_SC_SM.pdf}
\includepdf[pages=7]{PbTe_Tunable_SC_SM.pdf}
\includepdf[pages=8]{PbTe_Tunable_SC_SM.pdf}

\end{document}